\def \SAIT #1 #2 {{\em Mem.\ Soc.\ Astron.\ It.\/} {\bf #1}, #2}
\def \MESS #1 #2 {{\em The Messenger\/} {\bf #1}, #2}
\def \ASTRNACH #1 #2 {{\em Astron. Nach.\/} {\bf #1}, #2}
\def \AAP #1 #2 {{\em Astron. Astrophys.\/} {\bf #1}, #2}
\def \AAL #1 #2 {{\em Astron. Astrophys. Lett.\/} {\bf #1}, L#2}
\def \AAR #1 #2 {{\em Astron. Astrophys. Rev.\/} {\bf #1}, #2}
\def \AAS #1 #2 {{\em Astron. Astrophys. Suppl. Ser.\/} {\bf #1}, #2}
\def \AJ #1 #2 {{\em Astron. J.\/} {\bf #1}, #2}
\def \ANNREV #1 #2 {{\em Ann. Rev. Astron. Astrophys.\/} {\bf #1}, #2}
\def \APJ #1 #2 {{\em Astrophys. J.\/} {\bf #1}, #2}
\def \APJL #1 #2 {{\em Astrophys. J. Lett.\/} {\bf #1}, L#2}
\def \APJS #1 #2 {{\em Astrophys. J. Suppl.\/} {\bf #1}, #2}
\def \APSS #1 #2 {{\em Astrophys. Space Sci.\/} {\bf #1}, #2}
\def \ASR #1 #2 {{\em Adv. Space Res.\/} {\bf #1}, #2}
\def \BAIC #1 #2 {{\em Bull. Astron. Inst. Czechosl.\/} {\bf #1}, #2}
\def \JSQRT #1 #2 {{\em J. Quant. Spectrosc. Radiat. Transfer\/} {\bf
#1}, #2}
\def \MN #1 #2 {{\em Mon. Not. R. Astr. Soc.\/} {\bf #1}, #2}
\def \MEM #1 #2 {{\em Mem. R. Astr. Soc.\/} {\bf #1}, #2}
\def \PLR #1 #2 {{\em Phys. Lett. Rev.\/} {\bf #1}, #2}
\def \PASJ #1 #2 {{\em Publ. Astron. Soc. Japan\/} {\bf #1}, #2}
\def \PASP #1 #2 {{\em Publ. Astr. Soc. Pacific\/} {\bf #1}, #2}
\def \NAT #1 #2 {{\em Nature\/} {\bf #1}, #2}

\documentstyle{memsait}
\input epsf.sty
\begin{opening}
\title{INTERNAL SHOCKS IN THE JETS OF BLAZARS}
\author{M. Spada$^1$, D. Lazzati$^{2,3}$, G. Ghisellini$^3$, A.
Celotti$^4$}
\institute{$^1$ Osservatorio di Arcetri, Firenze, Italy\\
$^2$ Universit\`a di Milano, Milano, Italy\\
$^3$ Osservatorio di Brera-Merate, Merate (LC), Italy\\
$^4$ SISSA, Trieste, Italy}
\date{}
\end{opening}

\begin{document}

\oddpagefooter{}{}{} 
\evenpagefooter{}{}{} 
\ 
\bigskip

\begin{abstract} The development of instabilities leading to the 
formation of internal shocks is expected in the relativistic outflows
of both gamma--ray bursts and blazars.  The shocks heat the expanding
ejecta, generate a tangled magnetic field and accelerate leptons to
relativistic energies.  While this scenario has been largely
considered for the origin of the spectrum and the fast variability in
gamma--ray bursts, here we consider it in the contest of relativistic
jets of blazars.  We calculate the expected spectra, light curves and
time correlations between emission at different wavelengths.  The
dynamical evolution of the wind explains the minimum distance for
dissipation ($\sim 10^{17}$ cm) to avoid $\gamma$--$\gamma$ collisions
and the low radiative efficiency required to transport most of the
kinetic energy to the extended radio structures.  The internal shock
model allows to follow the evolution of changes, both dynamical and
radiative, along the entire jet, from the inner part, where the jet
becomes radiative and emits at high energies ($\gamma$--jet), to the
parsec scale, where the emission is mostly in the radio band
(radio--jet).

\end{abstract}

\section{Hierarchical internal shock model}

We propose that the {\it internal shock scenario},
which is the standard scenario
proposed to explain the observed gamma--ray burst radiation
(Rees \& M\'esz\'aros 1994, Lazzati et al. 1999, Panaitescu et al.
1999), can work also for radio sources in general, and for blazar 
in particular.
The central engine which gives origin to jets in radio sources
may work intermittently, accelerating shells of plasma with slightly
different mass, energy and velocity.
Faster and later shells can then catch up slower earlier ones.
In the resulting collisions a shock develops, converting some of
the ordered bulk kinetic energy into magnetic field and
random energy of the electrons which radiate.

The wind is treated as a sequence of $N=t_w/t_v$ shells, where $t_w$
is the duration time of the wind ejection from the central
source and $t_{v}\ll t_w$ is the average interval between consecutive
ejections. Each shell is characterized by a mass $M_j$, a Lorenz factor
$\Gamma_j$ and an ejection time $t_j$. After setting the dynamics of the 
wind ejection, we calculate the radii where the shells collide,
approximately
given by:
\begin{equation}
R=\frac{2\alpha^2}{\alpha^2-1}\Gamma_{\rm min}^2  c t_v ~~~~~
\alpha=\Gamma_{\rm max}/\Gamma_{\rm min}
\end{equation}
where $\Gamma_{\rm min}-\Gamma_{\rm max}$ is the range of bulk Lorentz 
factors of the shells. 
In the case of blazar jets $\Gamma\approx 10$ and $c t_v$ 
is assumed to be of the order of the dimension 
of the central engine $\approx 10^{14}$ cm (for a BH with 
$M\approx10^9 M_{\odot}$).
Thus the collisions start at $\sim 10^{16}-10^{17}$ cm, in agreement
with
the estimates of the source size given by the observed variability
and by the requirement of negligible $\gamma$--$\gamma$ collisions. 
The merged shells 
propagate on larger scales colliding again up to $ 10^{20}$ cm, 
where the differences among the shell velocities are
completely smoothed out and the wind can be considered uniform.

For each collision we study the hydrodynamics, in order to determine 
the shock velocity, the compression ratio and the internal energy 
$E_{sh}$ of the shocked fluid. 
Assuming a given partition of $E_{sh}$ among protons,
electrons ($E_e=\epsilon_e E_{sh}$, with $\epsilon_e\approx 0.5$) 
and magnetic field ($E_B=\epsilon_B E_{sh}$, $\epsilon_B\approx 0.04$),
we calculate the relevant physical parameters in the shocked fluid
and the emitted spectrum. 
The relativistic particles are assumed to have the same energy
distribution 
(a broken power--law) throughout the entire emitting zone. 
This simplification does not allow to consider details of
spectral changes on a timescale faster than the light crossing
time of a single shell (few hours when $R\approx 10^{17}$ cm
and a month on the parsec scale).

The relevant radiation processes are synchrotron,
synchrotron self--Compton (SSC), and Compton scattering 
on the external radiation (EC), likely produced by the broad line region (BLR).
The emission from each shocked region is assumed to last
for the longest among the following timescales: the cooling time 
of the electrons, the time for the shock to cross the two 
colliding shells, and the time taken by photons to escape the source.  
We simulate the evolution of the total spectrum summing the locally 
produced spectra of those regions of the jet which are simultaneously 
active in the frame of the observer.

\begin{figure}
\epsfxsize=11cm 
\hspace{2.cm}\epsfbox{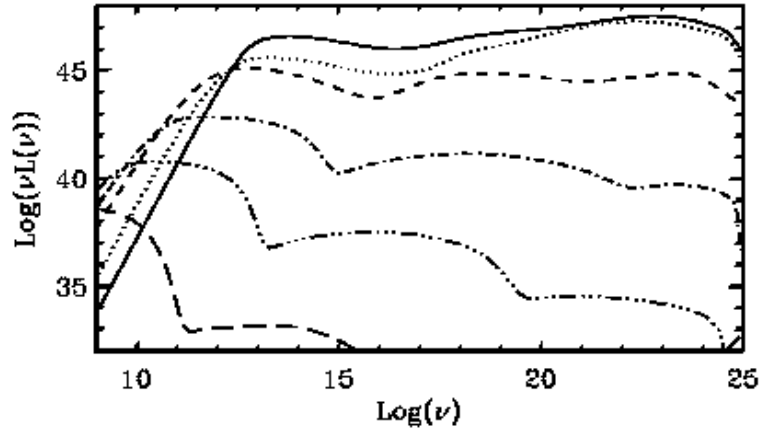} 
\caption[h]{ 
Average spectrum produced in different regions of the jet:
$R<10^{17}$ cm (solid line),
$10^{17}<R<5\times10^{17}$ cm (dotted line),
$5\times 10^{17}<R<2.5\times10^{18}$ cm (dashed line), 
$2.5\times 10^{18}<R<1.25\times10^{19}$ cm (dot-dashed line), 
$1.25\times10^{19}<R<6.25\times10^{19}$ cm (3dots-dashed line),
$R>6.25\times10^{19}$ cm (long-dashed line).
}
\end{figure}

\section{Results and Conclusions}

In order to compare the results of the internal shock model
with the observed spectrum and the temporal behavior of a powerful
blazar such as 3C~279, we simulated a wind with an average kinetic
luminosity $L_w\approx 10^{48}$ erg s$^{-1}$. 
The injection lasts for 10 years with an average interval between two 
consecutive shells of 3 hours (30000 shells in total). 
The shell Lorentz factors are random between $10<\Gamma_j<30$, 
the masses are also random with an average value of 
$M_j\approx 4\times 10^{-3} M_{\odot}$, and we require a constant 
luminosity for the wind. 
This implies that more energetic shells are followed by longer quiet
times, during which the central engine ``re-charges''.
The BLR has a radius of $R_{blr}=5\times 10^{17}$cm and 
a luminosity of $\sim 10^{45}$ erg s$^{-1}$. 

Since the Lorentz factors of the colliding shells are only slightly
different ($\Gamma_{\rm max}/\Gamma_{\rm min}=3$), the internal shock 
model is characterized by a low radiative efficiency: the fraction 
of kinetic energy converted into photons is less than 10\%. 

In Figure 1 we plot the average spectrum produced in different 
regions of the jet, in order to show the spectral evolution during
the wind expansion. 
For the inner radii, $R\le R_{blr}$, the external Compton emission 
dominates: most of the radiation is emitted at $10^{23}-10^{24}$Hz 
on time scales of the order of a few hours; 
the synchrotron spectrum peaks at  $\sim 10^{14}-10^{15}$ Hz, 
with a luminosity of $\approx 10^{46}$ erg s$^{-1}$.
At larger radii only the synchrotron and the SSC 
(first and second order) processes contribute. 
The shell--shell collisions become less efficient as the distance
increases,  
and the synchrotron peak shifts to lower energy, reaching $\sim$10 GHz 
for collisions at $10^{19}$ cm.  
The time scale of the emission increases from few hours to few weeks, 
and the synchrotron dominates over the inverse Compton power at 
large distances.

The main observed properties explained by the internal shock model are:
\begin{enumerate}
\item {\bf Low efficiency ---} The radiative output of radio sources
in
general and blazars in particular must be a small fraction (say less than
10\%) 
of the energy transported by the jet, since the extended radio
structures 
require a power input much exceeding what is lost through radiation.
\item {\bf Minimum distance for dissipation ---} The bulk of the power
we see must be produced at some distance ($> 10^{16}$ cm) from 
the jet apex and from the accretion disk.  
Otherwise the high energy $\gamma$--rays are absorbed
by a dense radiation field leading to electron--positron pairs and
to a softer (especially X--ray) radiation, which is not observed.
\item {\bf Continuous energy deposition along the jet --} 
Dissipation occurs all along the jet: the $\gamma$--ray flux is
produced mainly in the inner part, and can vary rapidly,
while the radio flux is produced by the parsec scale jet,
with a variability time scale of the order of weeks--months. 
\end{enumerate}

\end{document}